\documentclass[preprint,showpacs,preprintnumbers,amsmath,amssymb,
nofootinbib,eqsecnum,12pt,floatfix]{revtex4-1}

\usepackage{fullpage}
\usepackage{amsmath}
\usepackage{graphicx}
\usepackage{amssymb}

\usepackage{cancel}
\usepackage{amsmath,amsthm}

\numberwithin{equation}{section}

\newlength{\dummysp}
\newcommand{\tr}{\mathop{{\hbox{Tr} \, }}\nolimits}

\newcommand{\beq}{\begin{eqnarray}}
\newcommand{\eeq}{\end{eqnarray}}

\newcommand{\gappeq}{\mathrel{\rlap {\raise.5ex\hbox{$>$}}
{\lower.5ex\hbox{$\sim$}}}}
\newcommand{\lappeq}{\mathrel{\rlap{\raise.5ex\hbox{$<$}}
{\lower.5ex\hbox{$\sim$}}}}
\newcommand{\myref}[1]{(\ref{#1})}

\newcommand{\ben}{\begin{enumerate}}
\newcommand{\een}{\end{enumerate}}

\newcommand{\bit}{\begin{itemize}}
\newcommand{\eit}{\end{itemize}}

\newcommand{\Ncal}{{\cal N}}

\newcommand{\cA}{{\cal A}}

\newcommand{\cF}{{\cal F}}
\newcommand{\cFb}{{\overline{\cal F}}}
\newcommand{\cD}{{\cal D}}
\newcommand{\cDb}{{\overline{\cal D}}}
\newcommand{\cQ}{{\cal Q}}
\newcommand{\cU}{{\cal U}}

\newcommand{\cUb}{{\overline{\cal U}}}

\newcommand{\lambdalat}{\lambda_{\text{lat}}}
\newcommand{\slnc}{$SL(N,\mathbb{C})  $} 
\newcommand{\Tr}[1]{\ensuremath{\mbox{Tr}\left[ #1 \right]} }
\newcommand{\Ibb}{\ensuremath{\mathbb I} }

\def\[{\left [}
\def\]{\right ]}
\def\({\left (}
\def\){\right )}

\usepackage[colorlinks = true, linkcolor = blue ,urlcolor  = blue, citecolor = blue, anchorcolor = blue]{hyperref}

\begin{document}

\title{On the removal of the trace mode in lattice ${\cal N}=4$ super Yang-Mills theory}

\author{Simon Catterall}
\email{smcatter@syr.edu}
\affiliation{Department of Physics, Syracuse University, Syracuse, 13244 NY USA}
\author{Joel Giedt}
\email{giedtj@rpi.edu}
\affiliation{Department of Physics, Applied Physics and Astronomy,
Rensselaer Polytechnic Institute, 110 8th Street, Troy NY 12065 USA}
\author{Raghav G. Jha}
\email{rgjha@syr.edu}
\affiliation{Department of Physics, Syracuse University, Syracuse, 13244 NY USA}

\date{\today}

\begin{abstract}
Twisted and orbifold formulations of lattice ${\cal N}=4$ super Yang-Mills theory which possess an 
exact supersymmetry require a
$U(N)=SU(N)\otimes U(1)$ gauge
group. In the naive continuum limit, the $U(1)$ modes trivially decouple and play no role
in the theory. However, at non-zero lattice spacing they couple to the $SU(N)$ modes and
can drive instabilities in the lattice theory. For example, it is well known
that the lattice $U(1)$ theory undergoes a
phase transition at strong coupling to a chirally broken phase. An improved action that suppresses the
fluctuations in the $U(1)$ sector was proposed in \cite{Catterall:2015ira}. Here, we explore a more aggressive approach to
the problem by adding a term to the action which can entirely suppress the $U(1)$ mode. The penalty is that the
new term breaks the $\mathcal{Q}$-exact lattice supersymmetry. However, we argue that the term is $1/N^2$ suppressed and the
existence of a supersymmetric fixed point in the planar limit ensures that any SUSY-violating terms induced in
the action possess couplings that also vanish in this limit. We present numerical results on supersymmetric
Ward identities consistent with this conclusion.
\end{abstract}

\maketitle

\section{Introduction}
In recent years a great deal of effort has been devoted to the construction and numerical studies of
lattice formulations of $\Ncal=4$ super Yang-Mills theory which retain one exact supersymmetry at non-zero
lattice spacing---see the reviews \cite{Giedt:2006pd,Catterall:2009it,Bergner:2016sbv} and references therein.
These lattice theories 
can be derived using either deconstruction \cite{Cohen:2003xe, Cohen:2003qw, Kaplan:2005ta} or topological field theory methods 
\cite{Catterall:2003wd, Catterall:2005fd, Catterall:2007kn,Unsal:2006qp}. In this approach 
the link fields appearing in the lattice theory take their values in the
algebra of the group, denoted by $\mathfrak{gl}$($N,\mathbb{C})$.\footnote{This 
restriction is not present for Sugino's formulation---see \cite{Sugino:2003yb}.
Other approaches to studying ${\cal N}=4$ super Yang-Mills and the AdS/CFT 
correspondence on a
computer include \cite{Honda:2010nx,Hanada:2008ez,Anagnostopoulos:2007fw,Hanada:2010kt,
Hanada:2013rga,Honda:2011qk,Ishii:2008ib,Ishiki:2008te}.}
This is readily apparent from the (twisted) scalar supersymmetry (SUSY)
transformation
\beq
\cQ \cU_m = \psi_m
\label{impsusy}
\eeq
where $\psi_m$ is a twist fermion that transforms as a link variable.
Since it is a fermion, it has an expansion in terms of generators,
\beq
\psi_m = \sum_{A=0}^{N^2-1} \psi_m^A t^A
\eeq
Here, $t^0$ is proportional to the unit matrix, and must be included
if \myref{impsusy} is to hold, because the link field $\cU_m$ on the
left-hand side certainly has an expansion involving the unit matrix,
if it is to yield the usual $a \to 0$ continuum limit
\beq
\cU_m(x) = 1 + a \cA_m(x) + \cdots
\eeq
(Here, $\cA_m(x)$ is a complexification that contains both the
gauge fields and scalars.)
On the other hand, SUSY should not convert a group valued field into
a Lie algebra valued field, so in fact $\cU_m$ should also have the
expansion
\beq
\cU_m = \sum_{A=0}^{N^2-1} \cU_m^A t^A
\eeq
with the U(1) mode $\cU_m^0$ {\it fully dynamical.}
The conclusion of this argument is that the scalar SUSY $\cQ$
requires the gauge group to be U(N) and not SU(N), with the bosonic
link fields Lie algebra valued.

In the continuum
limit the entire U(1) sector decouples, and becomes an uninteresting free theory---all fields
are in the adjoint representation and hence neutral for U(1).  However on the lattice this sector is coupled
to the SU(N) part through irrelevant operators, so we cannot completely ignore it.  In fact, it is
these irrelevant couplings that can cause various problems. The first of these was first identified in \cite{Catterall:2014vka}
and is manifested in the appearance of a chirally broken phase for 't Hooft couplings $\lambdalat > 1$ (see
eqn.~\ref{susyaction} for the definition of the lattice coupling).

Another way to see that the $U(1)$ mode drives instabilities is to examine the behavior of the theory
under the classical
scaling transformation
\begin{eqnarray}
\cU_\mu &\to& c\cU_\mu\\\nonumber
\cUb_\mu &\to& c\cUb_\mu\\\nonumber
\psi_\mu &\to& c^{\frac{3}{2}}\psi_\mu\\\nonumber
\eta &\to& c^{\frac{3}{2}}\eta\\\nonumber
\chi_{\mu\nu} &\to& c^{\frac{3}{2}}\chi_{\mu\nu}
\label{scaling}
\end{eqnarray}
It is trivial to see that the supersymmetric action given
in \cite{Catterall:2014vka} (minus the soft $\cQ$-breaking
mass term) is invariant under this transformation if the Yang-Mills coupling $g^2\to c^4g^2$.
This allows us to write down relations between expectation values of gauge invariant operators. For example,
\beq
\langle {\rm Tr}\prod_{i=1}^P \cU^i \rangle_{g^2}
=c^P \langle {\rm Tr}\prod_{i=1}^P\cU^i \rangle_{c^4g^2}
\eeq
in which we have suppressed spacetime coordinates and indices and where $\cU$ could be replaced by
any other appropriately chosen lattice field with a corresponding change in the multiplicative factor on the RHS.
Since the LHS is independent of $c$ this implies that the expectation value on the RHS must vary as
$c^{-P}$. Note that
this rescaling is not allowed if the link variables $\cU_\mu$ are $SL(N,\mathbb{C})$-valued,
corresponding to gauge group SU(N).  Thus, it is the U(1) sector that creates this
instability.

In \cite{Catterall:2015ira} a new supersymmetric term was added to the lattice action to suppress the $U(1)$ mode
fluctuations. This allows for simulations to be performed out to stronger coupling $\lambdalat \leq 2$. However,
it does not appear sufficient to explore the regime of extreme strong coupling needed for
studies of S-duality \cite{Giedt:2016wig}. The reason for the ineffectiveness of this term at very strong coupling is
that it constrains only the real part of the determinant of the plaquette operator averaged over all plaquettes
associated with a given lattice site. 

In this paper, we have attempted to address this problem in a different way by adding to the lattice action
a term which explicitly suppresses the $U(1)$ sector for each link field (we call this the \textit{detlink} term). 
We argue that this term is 1/$N^2$ suppressed and hence the exact scalar SUSY $\cQ$ should be recovered in the large $N$ limit.
Furthermore, we show extensive numerical results that support this conclusion.
The existence of this supersymmetry at large $N$ then guarantees that under
renormalization any $\cQ$-breaking operators that are generated are $1/N^2$ suppressed,
and the scalar SUSY is restored without fine tuning as $N\to\infty$. In addition,
we show that even for modest values of $N$ such as $SU(5)$, $\cQ$ invariance
is a very good approximation.  Early results for this formulation have appeared in \cite{Giedt:2018fxe}. 

An alternate method to achieve the same 
result is by truncating the theory completely to gauge group SU($N$) 
by having links valued in the group $SL(N,\mathbb{C})$ rather than algebra $\mathfrak{gl}$($N,\mathbb{C})$. 
However, the full truncation (bosonic \& fermionic) 
of the theory from U($N$) to SU($N$) does not work. A simple way to see this is as follows: 
Assume a traceless fermion $\psi_a$ which lives on the link in the direction of $e_a$. 
The gauge invariance acts as 
\begin{equation}
\psi_a(x) \to G(x) \psi_a(x) G^\dagger(x + e_a),
\end{equation}
which yields a $\psi_a$ which is not in general traceless. 
Thus we cannot eliminate the $U(1)$ mode of the fermion, even
under the restriction to $SU(N)$ gauge group.\footnote{This is
to be contrasted with \cite{Berkowitz:2016jlq,Rinaldi:2017mjl,Berkowitz:2018qhn} where it was possible to eliminate
the $U(1)$ fermion mode.  This has the benefit of improving
the condition of the fermion matrix.}  Note that this is
a lattice effect, since for a site fermion $\eta$, we would have
\beq
\tr G(x) \eta(x) G^\dagger(x) = \eta^A(x) \tr G(x) t^A G^\dagger(x) = \eta^A \tr t^A = 0
\eeq
assuming $\eta(x)$ only involved the generators $t^A$ of $SU(N)$, which are traceless.
The distinction between link fermions and site fermions is only meaningful on the lattice.
This same argument does not apply to the link bosons, since they are valued in the group
and the gauge transformation preserves that feature.

In summary, to maintain lattice gauge invariance, for this {\it hybrid} action 
we only truncate the bosonic sector down to SU($N$). 
This construction also restores $\mathcal{Q}$ supersymmetry 
in the limit $N\to\infty$. In Table \ref{tab:ward-hybrid}, we show the comparison between these two approaches. This method of maintaining
exact lattice supersymmetry by truncating U(1) sector at large $N$ was employed in \cite{Catterall:2017lub,Jha:2017zad} 
to initiate non-perturbative checks of gauge/gravity duality at large $N$ in two dimensions. 
In this paper, we show detailed numerical results in four dimensions consistent with the claimed $1/N^2$ suppression.

\section{Lattice action}
The $\cQ$-exact lattice action takes the form
\begin{align}
  S & = \frac{N}{4\lambdalat} \sum_x \Tr{\cQ \left(\chi_{ab}\cD_a^{(+)}\cU_b + \eta \cDb_a^{(-)}\cU_a 
- \frac{1}{2}\eta d \right)} + S_{\text{cl}}  \\
  S_{\text{cl}} & = -\frac{N}{16\lambdalat} \sum_x \Tr{\epsilon_{abcde} \chi_{de}(x + e_a + e_b + e_c) 
\cDb^{(-)}_{c} \chi_{ab}(x + e_c)},
\label{susyaction}
\end{align}
where the lattice difference operators take the form of shifted commutators.
For example,
\begin{equation}
  \cD_a^{(+)} \cU_b(x) = \cU_a(x) \cU_b(x + e_a) - \cU_b(x) \cU_a(x + e_b) \equiv \cF_{ab}(x)
\end{equation}
where $e_a$ are the principle lattice vectors of the $A_4^*$ lattice.
The $\cQ$-closed term is still lattice supersymmetric due to the existence of an exact lattice Bianchi identity,
\begin{equation}
\epsilon_{abcde} \cDb^{(-)}_c \cFb_{ab}(x + e_c) = 0.
\end{equation}
After we integrate out the auxiliary field $d$, we have 
\beq
\label{eq:lat_act}
S  = \frac{N}{4\lambdalat} \sum_x \text{Tr} \left[ -\cFb_{ab} \cF_{ab} 
+ \frac{1}{2} \left( \cDb_a^{(-)} \cU_a\right )^2 
- \chi_{ab} \cD^{(+)}_{ [a } \psi_{ b] } 
- \eta \cDb^{(-)}_a \psi_a \right] + S_{\text{cl}}.
\eeq
The action also contains a single trace mass term, which helps to lift the classical flat directions by giving a small
mass to the scalar fields:\footnote{It also generates cubic and quartic terms that further stabilize the flat directions.
This mass term has been used for most of our earlier works, and also appears in \cite{Hanada:2010qg}.}
\beq
  \label{eq:single_trace}
  S_{\text{mass}} = \frac{N}{4\lambdalat} \mu^2 \sum_{x, a} \Tr {\bigg(\cU_a^\dagger \cU_a - \Ibb_N\bigg)^2}.
\eeq
To control the local fluctuations of the $U(1)$ sector we now add a new term to the action:
\beq
\Delta S = \frac{N}{4\lambdalat} \kappa_\text{link} \sum_{x,a} |\det \cU_a(x) - 1|^2
\label{det}
\eeq
In the limit $\kappa_\text{link}\to\infty$ we can completely remove the $U(1)$ modes --- both gauge and scalar
by restricting the links to \slnc. Notice that this term does {\it not} break the $SU(N)$ invariance of the
action since $\det \cU_a(x)$ is invariant under such transformations.
Using a polar decomposition of the link field
\beq
\cU_a(x)=(I+h_a)e^{iB_a}\eeq
the determinant can be written for small $h_a$ and $B_a$ as
\beq
{\rm det}\left(U_a\right)=(1+\frac{1}{\sqrt{N}}h_a^0)e^{i\frac{1}{\sqrt{N}}B_a^0}\eeq
where the $\frac{1}{\sqrt{N}}$ factor arises from the generators which satisfy the normalization ${\rm Tr}(T^aT^b)=-\delta^{ab}$
and the superscript indicates that only the trace mode survives.
To quadratic order in the fluctuations the determinant term becomes
\beq
\Delta S= \frac{1}{4\lambdalat} \kappa_\text{link} \sum_{x,a} \left(\left(B_a^0\right)^2+\left(h_a^0\right)^2\right)\eeq
The term thus serves to generate masses for the $U(1)$ modes. Additionally, notice it carries no prefactor of $N$ which then
guarantees that it will generate
terms that are $O(1/N^2)$ suppressed relative to the leading terms in a perturbative expansion.

This {\it detlink} term breaks both the $\cQ$-supersymmetry and the $U(1)$  gauge symmetry. Breaking the
$U(1)$ symmetry is likely harmless since the $U(1)$ sector plays no role in the continuum limit.
However breaking the exact supersymmetry is more problematic since it invalidates the arguments given in \cite{Catterall:2014mha}
devoted to the renormalizability of the lattice theory and specifically the number of counterterms needed to
tune to a supersymmetric continuum limit. 

To address this issue, we examine the $N$-dependence of the various terms in the action. It is clear that
the new term being a function of the trace modes only is suppressed by $1/N^2$ as compared to
all other terms in the action which correspond to a sum over all the generators of $U(N)$. 
If we treat this term perturbatively, it will yield a subleading contribution to
any observable in the planar limit. Thus, we expect that 
the exact supersymmetry will be restored in the large $N$ limit. The presence of an
exact supersymmetry at $N=\infty$ then ensures that any SUSY violating operators appearing at 
finite $N$ (and finite $\kappa_\text{link}$) are only multiplicatively renormalized with couplings 
proportional to positive powers of $1/N^2$.
In the next section, we show that these truncated approaches yield stable results 
for a range of values of the 't Hooft coupling $\lambdalat$ and measurements of appropriate Ward
identities show the expected $1/N^2$ behavior.

We perform the numerical simulations with the parallel code presented in \cite{Schaich:2014pda}. 
Since then, it has been extended to perform calculations for arbitrary gauge group to access the
large $N$ limit and will be presented in a future publication \cite{parallel_imp}.  We note that
there is an earlier work that develops a method to have $SU(N)$ gauge group in
supersymmetric lattice gauge theory \cite{Kanamori:2012et}.

\section{Ward identities}
\begin{figure}
\begin{center}
\includegraphics[width=5in]{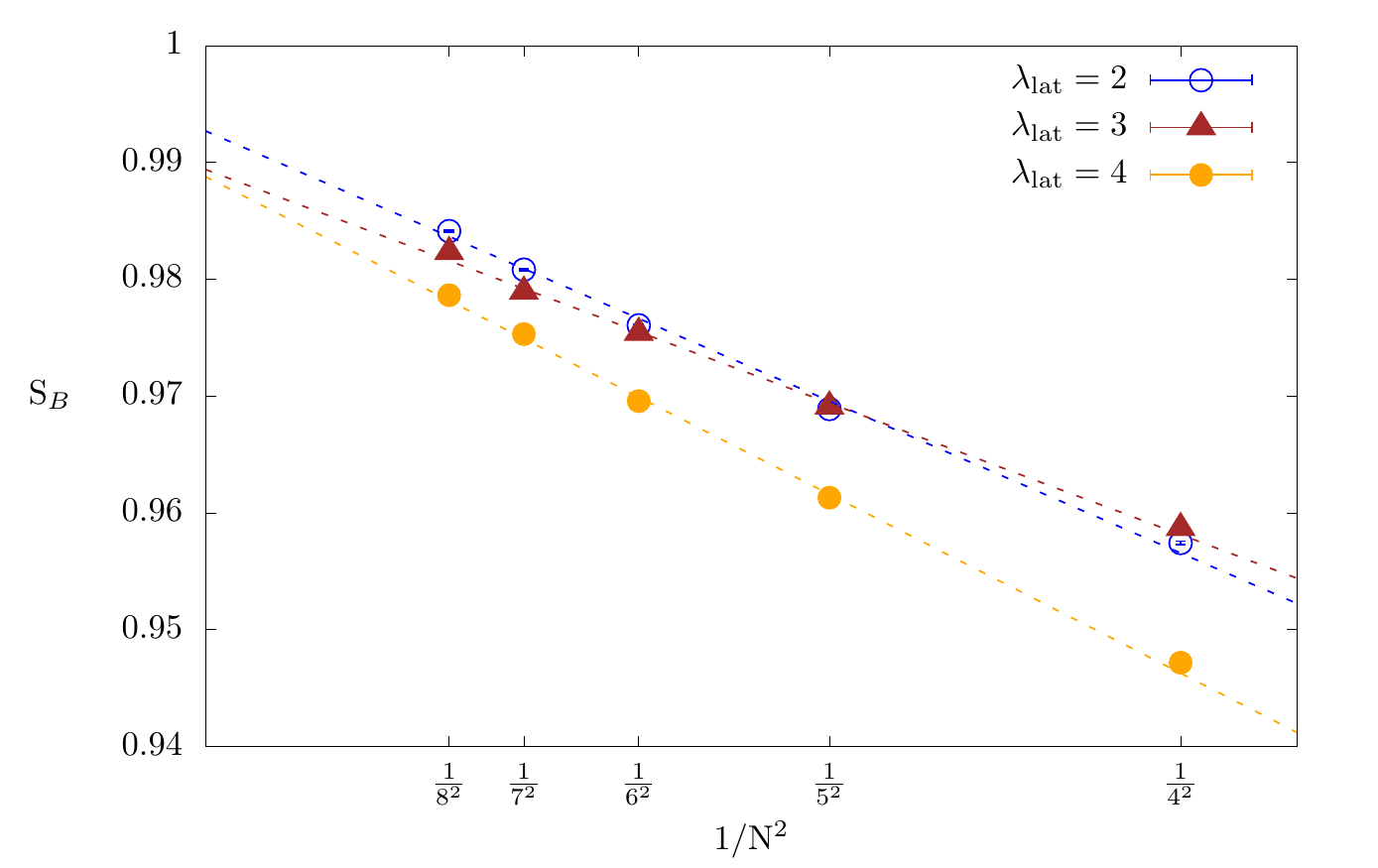}
\caption{The bosonic action, normalized such that it should be equal to $1$ if
the $\cQ$ symmetry is fully restored (exact).  It can be seen that the $N$ dependence
falls of as $1/N^2$, as expected.  The difference from $1$ in the large $N$
limit is anticipated from the presence of the small
mass term \myref{eq:single_trace} with $\mu=0.1$. Fits to $A + B/N^2$ are also shown in the plot.
For these runs we take $\kappa_\text{link}=5,5,10$ for the three values of $\lambdalat$ respectively.
\label{SBcombo}}
\end{center}
\end{figure}

We test the restoration of $\cQ$ in the large $N$ limit in two ways.\footnote{We note
that while $N=8$ is sufficient for us to see the large $N$ limit in our
four-dimensional lattices, much large $N$ are both necessary and
possible in the case of matrix quantum mechanics \cite{Berkowitz:2016jlq,Rinaldi:2017mjl,Berkowitz:2018qhn}.}  One is via
a measurement of the expectation value of the bosonic action $S_B$,
which is related to an exact lattice Ward identity associated with $\cQ$ in the
original, unmodified theory.
The results on $8^4$ lattice with three different values of $\lambdalat = 2, 3, 4$
are shown in Fig.~\ref{SBcombo}, with a normalization
such that $S_B = 1$ for exact $\cQ$.  It can be seen that the
restoration is within 1\% in the large $N$ limit, where presumably the
small deviation from $1$ is due to the mass term \myref{eq:single_trace} 
(we take $\mu = 0.1$ in our study) and thermal boundary conditions for the 
fermions along the temporal direction. 

\begin{figure}
\begin{center}
\includegraphics[width=5in]{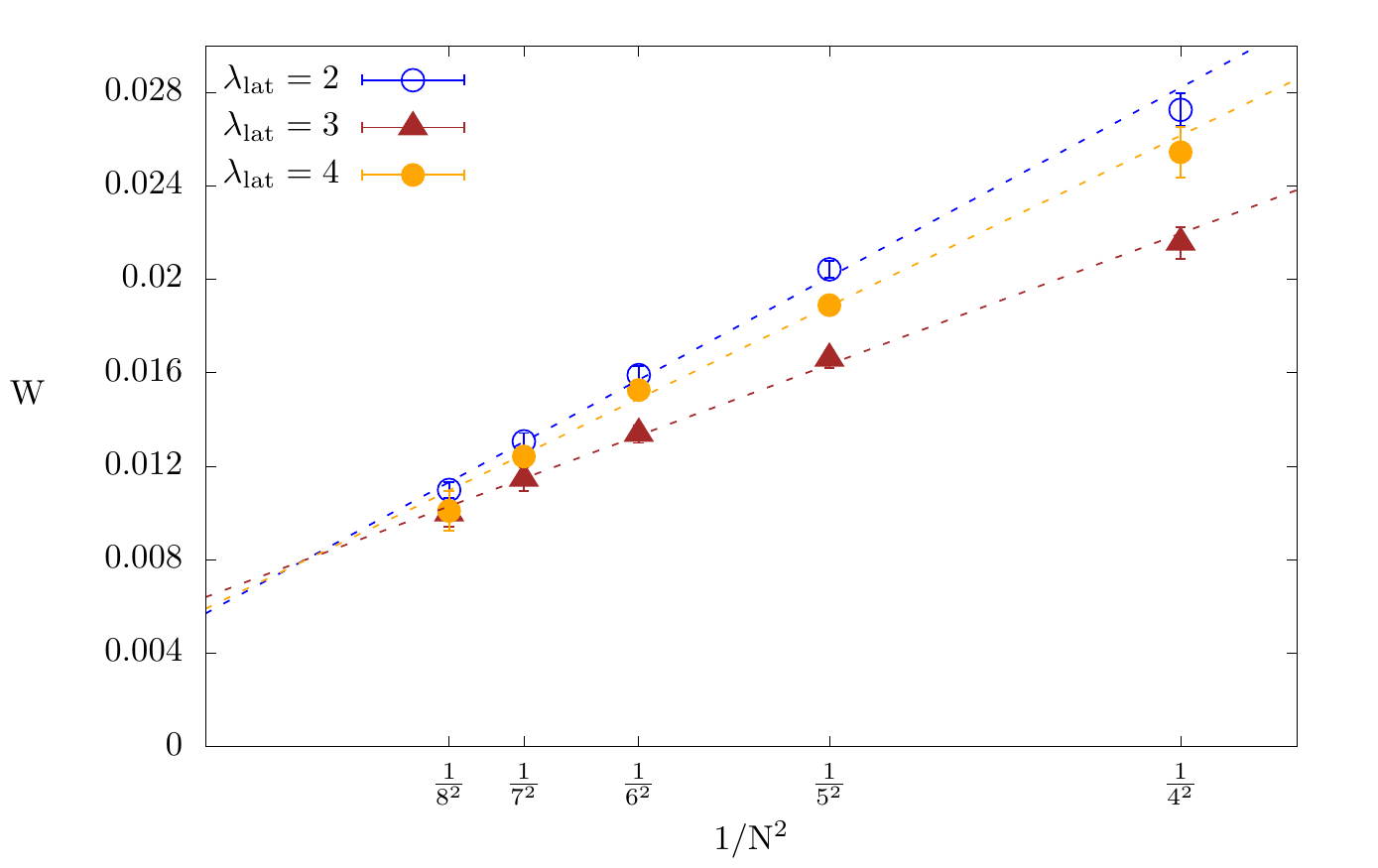}
\caption{The Ward identity \myref{wardid} for the $8^4$ lattice with detlink action, 
$\lambdalat=2,3,4$, $\mu=0.1$ and $\kappa_\text{link}=5,5,10$ respectively. 
Fits to $A + B/N^2$ are also shown. 
\label{wdl468}}
\end{center}
\end{figure}

Another check arises through the supersymmetric Ward identity corresponding to
\beq
\Big \langle  \cQ\, {\rm Tr}\,\left(\eta\cU_a\cUb_a\right) \Big \rangle=0\eeq
This yields
\beq
\Big \langle\,{\rm Tr}\, \left(d\cU_a\cUb_a\right)\Big \rangle - \Big \langle\,{\rm Tr}\,\eta \psi_a\cUb_a \Big \rangle=0\eeq
Using the equations of motion to eliminate $d$ we find
\beq
W= \Big \langle\,{\rm Tr}\,\left(\cDb_a \cU_a \cU_b\cUb_b\right)\Big \rangle - \Big \langle\,{\rm Tr}\,\eta \psi_a\cUb_a \Big \rangle=0\eeq
We further normalize $W$ by the fermion bilinear term appearing on the right and take the magnitude, 
\beq
W = \Bigg| \frac{\Big \langle\,{\rm Tr}\,\left(\cDb_a \cU_a \cU_b\cUb_b\right)\Big \rangle - \Big \langle\,{\rm Tr}\,\eta \psi_a\cUb_a \Big \rangle}{\Big \langle\,{\rm Tr}\,\eta \psi_a\cUb_a \Big \rangle} \Bigg|
\label{wardid}
\eeq
to obtain the quantity shown in Fig.~\ref{wdl468}.  It can be seen that the Ward identity, which is zero in the
limit of exact $\cQ$, is approximately $0.6$\% in the large $N$ limit.  Again, we attribute this to the
mass term \myref{eq:single_trace}.

\begin{figure}
\begin{center}
\includegraphics[width=5in]{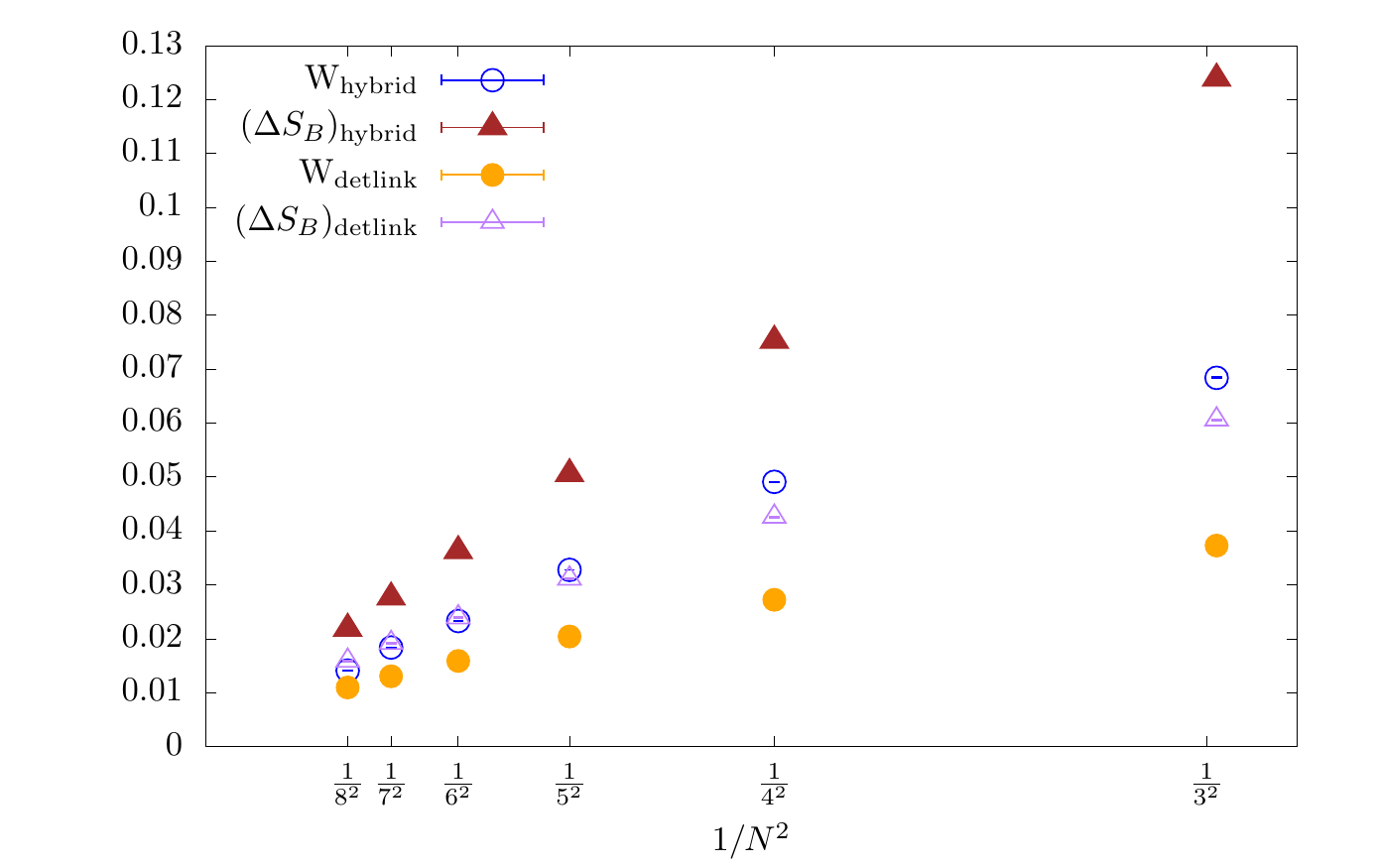}
\caption{\label{ward_hybrid}The comparison between the Ward identity results for the hybrid and detlink cases on $8^4$ lattice for $\lambdalat=2$. 
In the large $N$ limit, the difference is negligible.}
\end{center}
\end{figure}

We have also compared these results to the hybrid formulation, where the U(1) sector is
eliminated from the link fields entirely.  In Table \ref{tab:ward-hybrid} it can be seen
that the $\cQ$ violation is more for the \textit{hybrid} than in the \textit{detlink} formulation, but 
with the same 1/$N^2$ dependence. The results for the Ward
identity are shown together in Fig.~\ref{ward_hybrid}.  Thus we see that either approach will
restore $\cQ$ in the large $N$ limit, up to the effects of the regulating mass term.

\begin{figure}
\begin{center}
\includegraphics[width=5in]{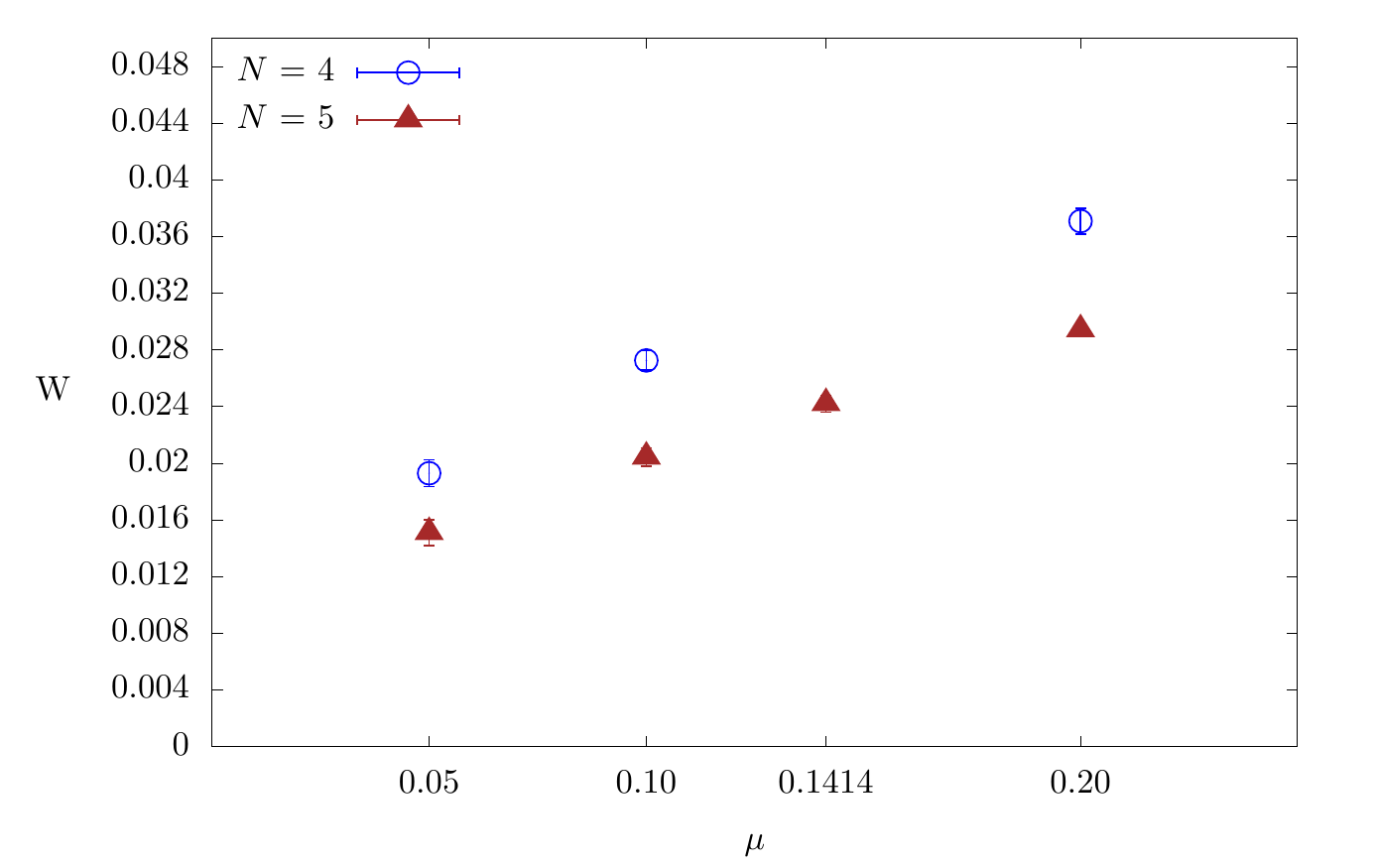}
\caption{Ward identity dependence on the mass parameter $\mu$. \label{mudep}}
\end{center}
\end{figure}

In Fig.~\ref{mudep} we show the dependence of the Ward identity on the mass term parameter $\mu$.  It can be seen that there is an appreciable decrease in the Ward identity as $\mu$ is decreased.  It is clear that our large $N$ extrapolation will also exhibit the same decrease with $\mu$ and it is reasonable to assume that the Ward identity will ultimately vanish as $N \to \infty$ and $\mu \to 0$.  For the latter limit it is important that the spacetime volume is also taken to infinity, since removing $\mu$ at finite $L$ will lead to unstable results as the scalar modes will wander without restriction.

A final question is the effect of finite volume, given that antiperiodic boundary
conditions are imposed on the fermions.  This also violates the $\cQ$ scalar supersymmetry,
so we expect such effects to fall off with the volume.  It can be seen from Table
\ref{tab:ward-hybrid2} that most of the volume effects are negligible.  Indeed, only
at the weakest coupling for the smallest number of colors is the effect of
any significance.

\begin{table}
\begin{center}  
\begin{tabular}{cccccccl} 
\hline\hline
$\lambdalat$ & $N$ & $\Delta S_{B}$ (detlink) & W (detlink) & $\Delta S_{B}$ (hybrid) & W (hybrid) \\
 \hline 
2.0 & 3 & 0.0606(1)  & 0.0373(8) & 0.1238(4)   & 0.0684(1) \\
& 4 & 0.0426(2)  & 0.0273(7) & 0.0753(2) &  0.0491(0)  \\
& 5 &  0.0311(1) & 0.0204(4)  & 0.0505(1)   & 0.0328(0)  \\
& 6 & 0.0239(1)  & 0.0159(4) &  0.0362(1)  & 0.0233(0)  \\
& 7 & 0.0192(1)  & 0.0131(4) &  0.0276(1)  & 0.0184(0)   \\
& 8 &  0.0159(1) & 0.0110(3) & 0.0218(1)  & 0.0141(0)  \\
\hline  \hline
\end{tabular}
\end{center}
\caption{
\label{tab:ward-hybrid} 
The comparison between the supersymmetry breaking observables using the detlink and the hybrid formulations
on $8^4$ lattice for $\lambdalat=2$. $\Delta S_{B}$ denotes the deviation from the supersymmetric value. See Fig.~\ref{ward_hybrid} for details.} 
\end{table}

\begin{table}
\begin{center}
\begin{tabular}{cccccccl} 
\hline\hline
 $\lambdalat$ & $N$ &  ~~~~~~~~~~~~~~~~~~~~~~~$8^{4}$  &  & ~~~~~~~~~~~~~~~~~~~~~~~$16^{4}$  \\ 
\cline{1-2}
 & & $\Delta S_{B}$ & W  & $\Delta S_{B}$  & W  \\
 \hline 
2.0 & 3 & 0.0606(1)   &  0.0373(8)  & 0.0407(18)   & 0.207(19) \\
& 4 & 0.0426(2) & 0.0273(8) & 0.0425(0) &  0.0281(2)  \\
& 5 &  0.0311(1) &  0.0204(8) & 0.0310(1)   & 0.0202(3)  \\
\hline
3.0 & 4 & 0.0413(2) & 0.0216(7) & 0.0420(2)  & 0.0218(3) \\
& 5 & 0.0309(1)  & 0.0166(4) & 0.0336(1) &  0.0174(4)  \\
\hline
4.0 & 3 & 0.0781(3)  & 0.0336(14) & 0.0788(1)   & 0.0357(3) \\
& 4 & 0.0528(2)  & 0.0254(11) & 0.0521(1) &  0.0230(4)  \\
\hline  \hline
\end{tabular}
\end{center}
\caption{
\label{tab:ward-hybrid2} 
The comparison between supersymmetry breaking observables using the detlink code on $8^4$ and $16^4$ lattices. The volume effects are small
in comparison at fixed $N$.} 
\end{table}

\section{Conclusions}
We have shown that simulations of lattice ${\cal N}=4$ SYM targeting the 
$SU(N)$ rather than the $U(N)$ theory are possible at moderately strong coupling $\lambdalat\le 4$. This
is a stronger coupling than has been achieved with the improved action described in \cite{Catterall:2015ira},
where only $\lambdalat \le 3$ was possible. 
In the case of gauge group $SU(2)$ simulations have even been performed at $\lambdalat=6$. However, unfortunately so far, 
we have not been able to extend this to even stronger couplings. Instead we observe the system appears to undergo a crossover
or phase transition to a regime in which the fermion operator develops very many small eigenvalues. We attribute this
to the presence of residual supersymmetry breaking associated with the determinant term. Work is underway to
develop a supersymmetric link based determinant term which may allow us to bypass these
problems and access yet stronger couplings.  The improvement that we do see is reflective of control
over the instabilities associated with the flat direction exhibited in the scaling \myref{scaling}.
The corresponding U(1) fluctuations are much more dangerous than the SU(N) related
flat directions because they allow the theory to wander into regimes associated with coarser
lattice spacings, where confinement is a generic feature.  In the future, we will
present results where further improvements can be obtained by preserving $\cQ$ exactly
while still controlling this U(1) sector in a rather aggressive way.

\section*{Acknowledgements}
SC and RGJ were supported by the US Department of Energy (DOE), Office of Science, 
Office of High Energy Physics, under Award Number DE-SC0009998. JG was supported in part by the Department of Energy,
Office of Science, Office of High Energy Physics,
Grant No. DE-SC0013496. Numerical calculations were carried out on the Center for Computational Innovations at Rensselaer 
and DOE-funded USQCD facilities at Fermilab.  We thank David Schaich for comments and discussions.

\bibliography{truncation}
\bibliographystyle{apsrev4-1}

\end{document}